\newcommand{\bkappa}{\mbox{\boldmath $\kappa$}}

\newcommand{\br}{\mbox{\boldmath $r$}}

\newcommand{\bk}{\mbox{\boldmath $k$}}

\RequirePackage{fix-cm}
\documentclass[twocolumn,epjc3]{svjour3}  
\smartqed  
\RequirePackage{graphicx}
\usepackage[numbers]{natbib}
\journalname{XXX}
\begin{document}

\title{Unintegrated gluon distributions from the color dipole 
cross section in the BGK saturation model 
}


\author{Agnieszka \L uszczak\thanksref{e1,addr1}
        \and
        Marta \L uszczak \thanksref{e2,addr2} 
        \and
        Wolfgang Sch\"afer
     \thanksref{e3,addr3} 
}

\thankstext{e1}{e-mail: Agnieszka.Luszczak@pk.edu.pl}
\thankstext{e2}{e-mail: mluszczak@ur.edu.pl}
\thankstext{e3}{e-mail: Wolfgang.Schafer@ifj.edu.pl}

\institute{Cracow University of Technology, Institute of Physics, PL-30-084 Krak\'ow, Poland \label{addr1}
           \and
           College of Natural Sciences, Institute of Physics, University of Rzeszów,
ul. Pigonia 1, PL-35-959 Rzeszów, Poland 
          \label{addr2}
           \and
           Institute of Nuclear Physics Polish Academy of Sciences, PL-31-342 Krak\'ow, Poland
           \label{addr3}
}

\date{Received: date / Accepted: date}

\maketitle

\begin{abstract}
We evaluate the unintegrated gluon distribution of the proton starting from a parametrization of the color dipole cross section including DGLAP evolution and saturation effects. To this end, we perform the Fourier-Bessel transform of $\sigma(x,r)/\alpha(r)$. At large transverse momentum of gluons we match the so-obtained distribution to the logarithmic derivative of the collinear gluon distribution.
We check our approach by calculating the proton structure function $F_L(x,Q^2)$ finding good agreement with HERA data.
\end{abstract}

\section{Introduction}
\label{intro}

The color dipole approach provides a useful basis for a phenomenology of high-energy processses in QCD.
On the one hand, the interaction of small size color dipoles with nucleons is governed by perturbative QCD (pQCD). Here the dipole cross section is linked to the gluon distribution of the target proton \cite{Nikolaev:1994ce}.
On the other hand, by insisting that the basic dipole-factorization holds also at large dipole sizes, the approach can be extended also to soft high-energy processes. This flexibility is one of the main advantages of the approach.

Many observables, such as deep inelastic scattering (DIS) structure functions, cross sections for diffractive DIS or exclusive vector meson production in
$\gamma^* p \to V p$ reactions can be straightforwardly evaluated directly in the color dipole picture.

However for many observables, such as jet production, dijet angular correlations or inclusive partivle production in hadronic collisions, a transition to momentum space is more convenient, see for example \cite{Nikolaev:2004cu}.
Here the color-dipole cross section will be replaced by the unintegrated gluon distribution (UGD), and the approach merges with the $k_T$-factorization \cite{Catani:1990eg,Levin:1991ry,Collins:1991ty} for small-$x$ processes.

In this letter we wish to present unintegrated gluon distributions  obtained from a dipole cross section fit \cite{Luszczak:2013rxa,Luszczak:2016bxd} which gives a very good description of precise HERA data on proton structure functions.

The direct modelling of the UGD in momentum space appears to be rather challenging in comparison, especially concerning the nonperturbative region of small momenta. For 
some recent approaches see for example \cite{Ermolaev:2017qei,Bacchetta:2020vty}.
\section{Unintegrated gluon distributions from the dipole cross section}
In the BGK approach \cite{Bartels:2002cj}, the color dipole cross section has the form
\begin{eqnarray}
\sigma(x,r) = \sigma_0 \Big( 1 - \exp\Big[-{\sigma_{\rm DGLAP}(x,r)\over \sigma_0} \Big] \Big) \, ,
\label{eq:BGK}
\end{eqnarray}
where for small dipole sizes, $\sigma(x,r)$ will be proportional to the 
to the DGLAP-evolving collinear gluon distribution of the proton, $\sigma(x,r) \to \sigma_{\rm DGLAP}(x,r)$: 
\begin{eqnarray}
\sigma_{\rm DGLAP}(x,r) = {\pi^2 \over N_c} \alpha(r) r^2   x g_{\rm DGLAP}\Big(x,{C \over r^2} + \mu_0^2 \Big) \, , 
\end{eqnarray}
with $N_c = 3$, and
\begin{eqnarray}
 \alpha(r) \equiv \alpha_s\Big({C \over r^2} + \mu_0^2 \Big)  \, .
\end{eqnarray}
For large dipole sizes, the dipole cross section approaches a constant
\begin{eqnarray}
\sigma(x,r) \to \sigma_0 \, , 
\end{eqnarray} 
while the strong coupling freezes at large $r$:
\begin{eqnarray}
\alpha(r) \to \alpha_{\rm fr} = \alpha_s(\mu_0^2) \, .
\end{eqnarray}
The behaviour in the infrared is thus governed by $\sigma_0$ and $\mu_0$ which are parameters to be fitted to an appropriate experimental observable. The dimensionless number $C$ is fixed as $C=4$.
\begin{table}
\caption{BGK fit  from Table 6 of Ref.~\cite{Luszczak:2016bxd} for $\sigma_r$  for H1ZEUS-NC data in the range $Q^2 \ge 3.5$~GeV$^2$ and $x\le 0.01$.NLO fit. { \it Soft gluon}.   $ m_{uds}= 0.14, m_{c}=1.3$ GeV. $\mu_0^2=1.9$ GeV$^2$.}
\begin{center}
\begin{tabular}{|c|c|c|c|c|} 
\hline 
$\sigma_0 [mb]$ & $A_g$ & $\lambda_g$ & $C_g$ &  $\chi^2/N_{df}$\\
\hline
 105.20$\pm$ &  2.4788$\pm$ & -0.066$\pm$ &  6.9093 $\pm$ & 1.04 
 \\
  12.234 & 0.093&  0.004&  0.510 &  \\
\hline
\end{tabular}
\end{center}
\end{table}
The gluon density, which  is parametrized  at the starting scale $\mu_{0}^{2}$, 
is evolved to larger scales, $\mu^2$ using NLO DGLAP evolution.
 We consider here the following form of the gluon density: 
\begin{equation}
   xg(x,\mu^{2}_{0}) = A_{g} x^{-\lambda_{g}}(1-x)^{C_{g}},
\label{gden-soft}
\end{equation}
the {\it soft} ansatz, as used in the original BGK model: 
The free parameters for this model are $\sigma_{0}$ and the  parameters for gluon $A_{g}$, $\lambda_{g}$, $C_{g}$. Their values are obtained by a fit to the data. It is also possible to vary the parameter $\mu^{2}_{0}$. However, to assure that the evolution is  performed in the perturbative region and to be compatible with the standard pdf fits we took  as a starting scale  $\mu^2_0 = 1.9$.  In the BGK model, the  $\mu_0^2$ scale is the same as the  $Q_0^2$ scale of the standard QCD pdf fits. \\
The color dipole cross section is related to the UGD of the target as \cite{Nikolaev:1994ce}
\begin{eqnarray}
\sigma(x,r) &=& {4 \pi \over N_c} \,  \alpha(r) \int {d^2 \bkappa \over \bkappa^4} {\cal F}(x,\bkappa) \Big[ 1 - \exp(i \bkappa \br) \Big] \, \nonumber \\
&=& {\pi^2 \over N_c} \alpha(r) r^2 \int_0^\infty {d\bkappa^2 \over  \bkappa^2} {4 [ 1- J_0(|\bkappa|r)] \over \bkappa^2 r^2} {\cal F}(x,\bkappa)
\label{eq:dipole}
\end{eqnarray}
For large dipole sizes, it approaches
\begin{eqnarray}
\sigma_0 = {4 \pi \over N_c} \alpha_{\rm fr} \int {d^2 \bkappa \over \bkappa^4}  {\cal F}(x,\bkappa) \, .
\end{eqnarray}
Before we proceed to inverting the Fourier transform (FT) of eq.~\ref{eq:dipole}, let us introduce for convenience
\begin{eqnarray}
f(x,\bkappa) \equiv {4 \pi \over N_c}  {{\cal F}(x,\bkappa) \over \bkappa^4} \, .
\end{eqnarray}
Notice, that often in the literature eq.~\ref{eq:dipole} is written with the strong coupling under the transverse momentum integral, so that in effect the FT maps $\sigma(x,r) \leftrightarrow \alpha_s {\cal F}(x,\bkappa)$.
However, if the UGD is to be used in a variety of hard processes beyond DIS, it would seem odd that it should have the factorization scale encoded. Moreover, as has been noted by Giraud and Peschanski  \cite{Giraud:2016lgg} , the nonanalytic behaviour $\propto 1/\log r$ spoils the positivity of the UGD obtained by inverting the FT.
As a side remark, we note, that the popular Golec-Biernat and W\"usthoff fit of the dipole cross section of Ref.~\cite{Golec-Biernat:1998zce} in fact assumes a fixed $\alpha_s \sim 0.2$.
We therefore propose to perform the Fourier invasion not of the dipole cross section, but of $\sigma(x,r)/\alpha(r)$.  
Then, by performing the FT of Eq.\ref{eq:dipole}, we obtain
\begin{eqnarray}
\int d^2 \br \exp[-i \bk \br] \, {\sigma(x,r) \over \alpha(r)} &=& (2 \pi)^2 \delta^{(2)}(\bk) \int d^2 \bkappa f(x,\bkappa) \nonumber \\
&-& (2 \pi)^2 f(x,\bk) \, ,
\end{eqnarray} 
or, alternatively
\begin{eqnarray}
f(x,\bk) = {\sigma_0 \over \alpha_{\rm fr}} \int {d^2 \br \over (2 \pi)^2}
e^{-i \bk \br} \Big( 1 - {\sigma(x,r) \over \sigma_0}{\alpha_{\rm fr} \over \alpha(r) }  \Big) \, .  
\end{eqnarray}
We can finally write this for the unintegrated glue ${\cal F}(x,\bk)$ as 
\begin{eqnarray}
{\cal F}(x,\bk)  &=& \bk^4 {\sigma_0 \over \alpha_{\rm fr}} {N_c \over 8 \pi^2} \nonumber \\
&\times& \int_0^\infty  r dr \,  J_0( |\bk|r) \Big( 1 - {\sigma(x,r) \over \sigma_0}{\alpha_{\rm fr} \over \alpha(r) }  \Big) \, .  
\label{eq:uglu}
\end{eqnarray}
The so defined unintegrated glue depends only on the gluon longitudinal momentum fraction and transverse momentum. This is typical of a small-$x$ approach.
Although our prescription does not catch all the powers of $\alpha(r)$
in the BGK parametrization of eq.~\ref{eq:BGK}, it still helps to get a more benign behaviour of the FT as far as unphysical numerical oscillations at large $\bk^2$ are concerned. In order to be able to extend the UGD to large $\bk^2$, we match the FT with the logarithmic derivative of the collinear glue of the BGK fit:
\begin{eqnarray}
{\cal F}_{\rm DGLAP}(x,\bk) = {\partial x g_{\rm DGLAP}(x,\bk^2) \over \partial \log \bk^2} \, .
\label{eq:DGLAP}
\end{eqnarray}
We find, that by choosing $\bk^2_{\rm match} = 20 \, \rm{GeV}^2$, we can find an excellent reproduction of the original $\sigma(x,r)$ when inserting the so-matched UGD back into eq.~\ref{eq:dipole}.
\section{Numerical results}
\begin{figure}
  \includegraphics[width=.5\textwidth]{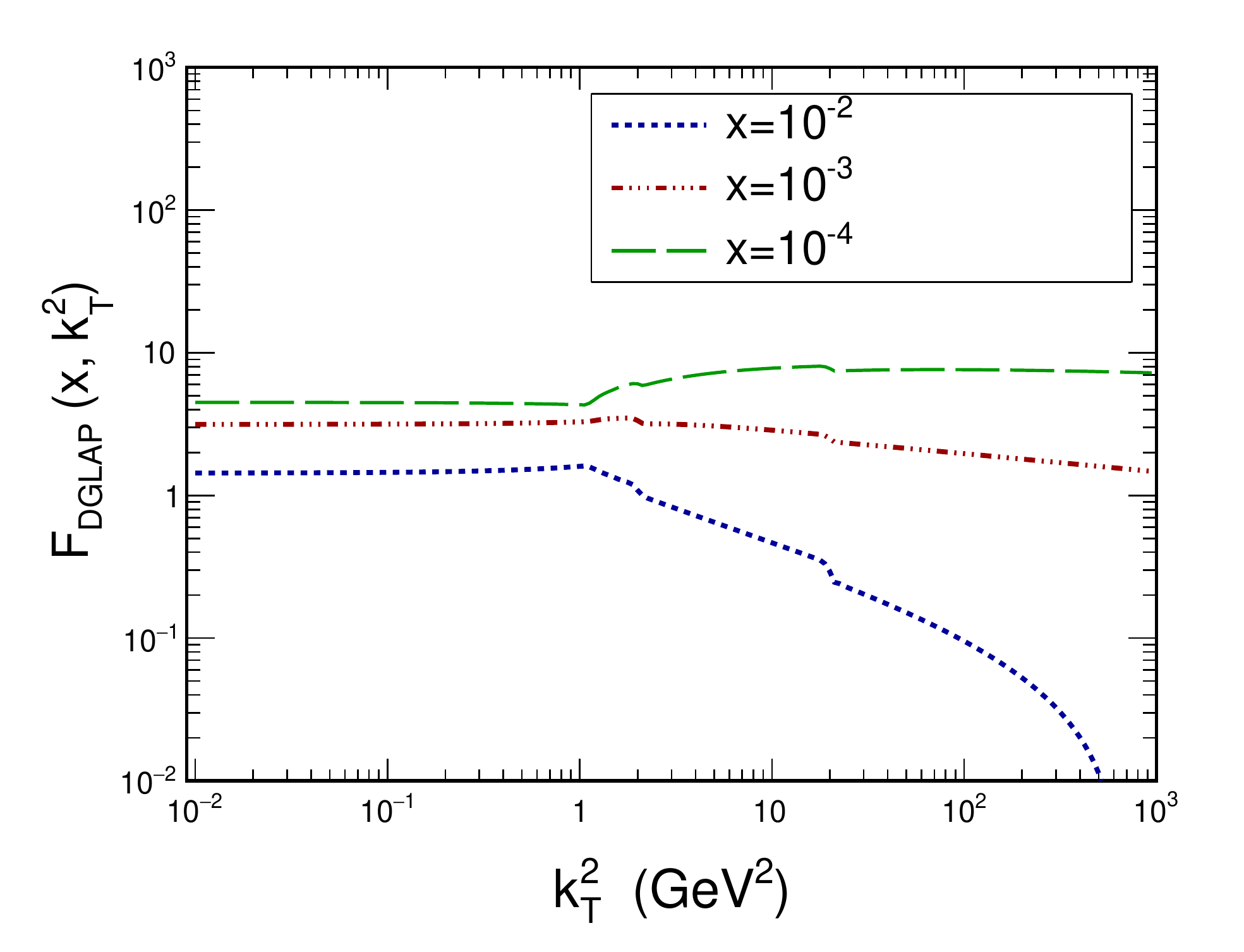}
\caption{Unintegrated glue in the DGLAP approximation from eq.~(\ref{eq:DGLAP}).}
\label{fig:1}       
\end{figure}
\begin{figure}
  \includegraphics[width=0.5\textwidth]{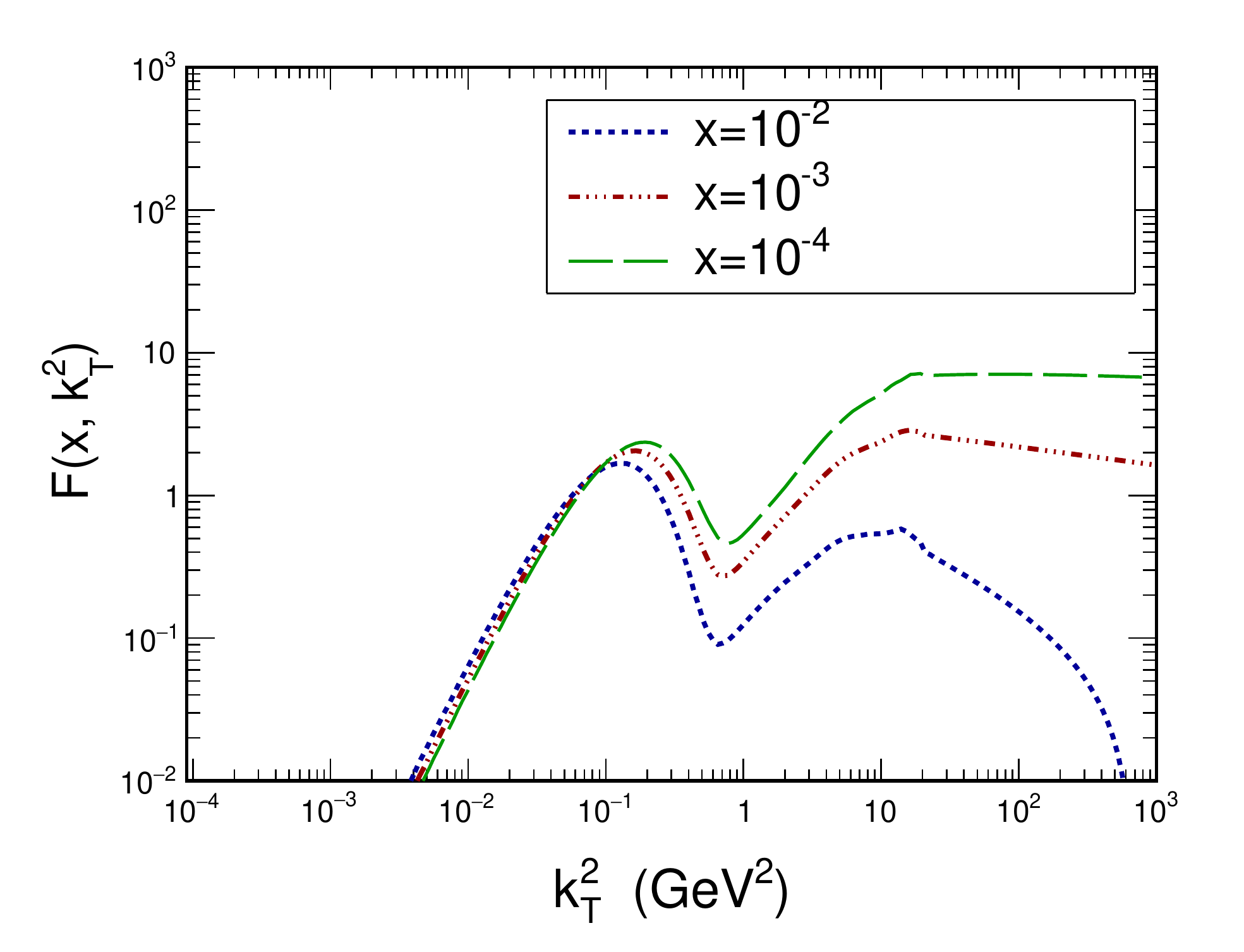}
\caption{Unintegrated glue from the Fourier transform of eq.~(\ref{eq:uglu}) merged with the DGLAP UGD at $\bk^2_{\rm match} = 20 \, \rm{GeV}^2$.}
\label{fig:2}       
\end{figure}
\subsection{Unintegrated glue}
In Fig.~\ref{fig:1} we show the logarithmic $\bk^2$-derivative of the input collinear glue as a function of $\bk^2$ for three values of $x$. We observe a hard tail at $\bk^2 > 1 \, \rm{GeV}^2$ which is rapidly increasing at small $x$. There is also a substantial $x$-dependence of the soft plateau at $\bk^2 < 1 \, \rm{ GeV}^2$.

Our main result, the UGD related to the BGK dipole cross section is shown in Fig.~\ref{fig:2}. A quite peculiar shape emerges, with a pronounced bump in the soft region and a local mimimum in the $\bk^2 \sim 1 \, \rm{GeV}^2$ region.
While the fall-off for $\bk^2 \to 0$ has only a very weak $x$-dependence, the latter is stronger in the region between the local maximum and minimum in the soft region. at perturbatively large $\bk^2$, there is a substantial low-$x$ growth of the UGD. The tail at $\bk^2 > \bk^2_{\rm match} = 20 \, \rm{GeV}^2$ finally is inherented from the log-derivative of the collinear glue.
It would be interesting to see, if the peculiar shape for low to intermediate $\bk^2$ can have observable consequences. Here especially the diffractive production of light vector mesons might be of interest, see e.g. \cite{Bolognino:2019pba,Bolognino:2021niq,Cisek:2022yjj}.
We also note, that depending on the observable, the region of small $\bk^2$ may be affected by Sudakov-type suppression. We also investigated another fit from Ref.~\cite{Luszczak:2016bxd}, finding very similar results.
\subsection{Proton structure function $F_L(x,Q^2)$}
\begin{figure}
  \includegraphics[width=0.5\textwidth]{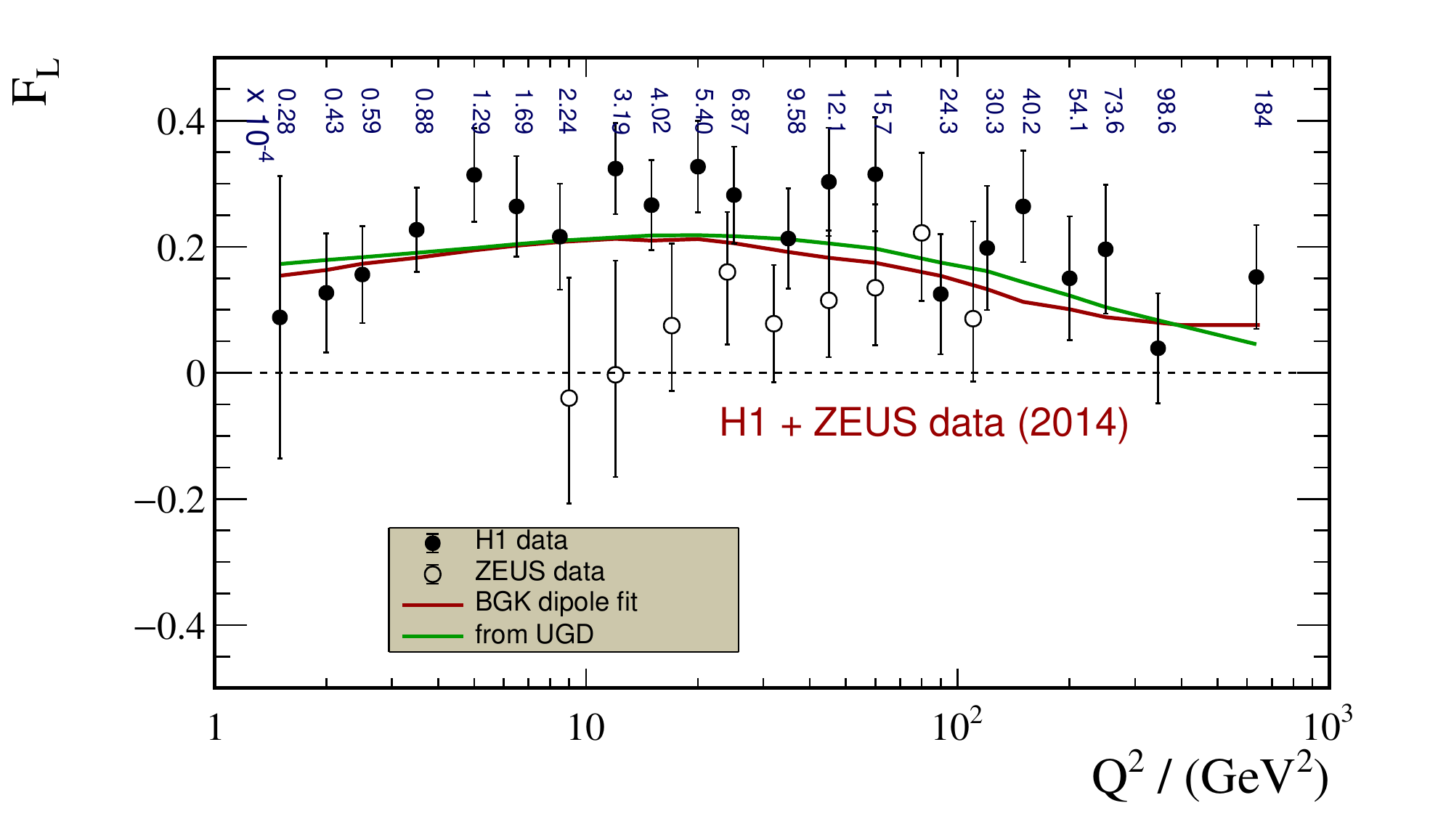}
\caption{Our results for the structure function $F_L(x,Q^2)$ using the dipole approach (red curve) and the $k_T$-factorization (green curve). Data are from Refs.~\cite{H1:2013ktq,ZEUS:2014thn}.}
\label{fig:3}       
\end{figure}
The longitudinal structure function at large $Q^2$ is a well-known probe of the gluon distribution of the proton \cite{Roberts:1990ww}. 
Data on $F_L$ is much more scarce than for $F_2$, for which a large body of measurements in a wide kinematic range is available.
Here descriptions based on the dipole approach like ours have the advantage that they can be extended also into the region of small $Q^2$, where standard collinear QCD-factorization is unreliable. Also, (a class of) higher twist effects are encoded in the dipole cross section, and hence also in the UGD obtained by us.
Therefore approaches like the one adopted here can be useful in making predictions e.g. for future experiments at an electron-ion collider. For a closely related modeling of $F_L$, see \cite{Badelek:2022cgr}. 
In Fig.~\ref{fig:3} we show our results for $F_L(x,Q^2)$ compared to the HERA data of the H1 and ZEUS collaborations \cite{H1:2013ktq,ZEUS:2014thn}. We show numerical results following two methodologies.
The longitudinal structure function $F_L$ is related to the virtual photoabsorption cross section for longitudinally polarized photons as
\begin{eqnarray}
F_L(x,Q^2) = {Q^2 \over 4 \pi^2 \alpha_{\rm em}} \, \sigma_L(x,Q^2) \, .
\end{eqnarray}
We evaluate $\sigma_L(x,Q^2)$ in two different methodologies.
Firstly, from the dipole approach
\begin{eqnarray}
\sigma_L(x,Q^2) = \sum_{f \bar f} \int_0^1 dz \int d^2\br \, \Big|\Psi_{\gamma^*}^{(f \bar f)}(z,\br,Q^2) \Big|^2 \, \sigma(x,r) \, , \nonumber \\
\end{eqnarray}
and secondly, from the $k_T$-factorization approach, where we have (see e.g. \cite{Ivanov:2000cm} for more details):
\begin{eqnarray}
\sigma_L(x,Q^2) &=& {\alpha_{\rm em} \over \pi} \int_0^1 dz \int d^2 \bk \int {d^2 \bkappa \over \bkappa^4} \alpha_s(q^2) {\cal{F}}(x_g,\bkappa) \nonumber \\
&\times& 4 Q^2 z^2(1-z)^2 \Big(
{ 1 \over \bk^2 + \varepsilon^2} - {1 \over (\bk-\bkappa)^2 + \varepsilon^2} \Big)^2 \, . \nonumber \\
\end{eqnarray}
Here $\varepsilon^2 = m_f^2 + z(1-z)Q^2$, and
we choose the renormalization scale as $q^2 = \max\{ \bk^2 + \varepsilon^2, \bkappa^2 \}$.
In Fig.~\ref{fig:3} we show our results from both approaches. We observe that both are consistent with each other. The description of data is good, also in the region of low $Q^2 \sim$ a few GeV$^2$.
\section{Summary and Outlook}
\label{sec:1}
In this letter, we have presented an unintegrated gluon distribution 
obtained from a BGK-type fit of the color dipole cross section to HERA structure function data. We argued that the Fourier transform to momentum space is best performed on $\sigma(x,r)/\alpha(r)$ instead of the dipole cross section itself.
The so-obtained UGD was then matched to a perturbative tail calculated from the DGLAP-evolved collinear glue that enters the BGK construction. 
We have obtained a good agreement of the proton structure function $F_L$, using both the color dipole factorization and $k_T$-factorization.
The strength of the approach lies in its ability to effectively include higher-twist corrections as well as its applicability in the soft region. A caveat concerns the precision of the pQCD part. Here more theoretical work is necessary. Firstly, as common in approaches derived from the small-$x$ limit, our UGD carries no dependence on the hard scale of the process.
To introduce this dependence,
one may follow a procedure proposed in Ref.~\cite{Kutak:2014wga}.
Secondly, while the collinear glue 
in the BGK ansatz is evolved by NLO DGLAP equations, our factorization formulas are at LO only. Here it may be argued that the error of this mismatch is smaller than it would be if NLO effects in the evolution were neglected altogther.
Finally also issues regarding the heavy quark scheme need to be clarified if one is concerned about precision predictions in the pQCD domain for future experiments. 
An interesting future extension would be to study the impact parameter dependent gluon distribution.
%
%
%
\begin{acknowledgements}
The work of A.L. was partially supported by NCN, Miniatura DEC-2021/05/X/ST2/00946.
\end{acknowledgements}

\bibliographystyle{apsrev.bst}
\bibliography{biblio}   


\end{document}